# Service Oriented Paradigm for Massive Multiplayer Online Games

Farrukh Arslan

*ECE Department, Purdue University, IN, USA*
*farslan@purdue.edu*

*Abstract.* In recent times Massive Multiplayer Online Game has appeared as a computer game that enables hundreds of players from all parts of the world to interact in a game world (common platform) at the same time instance. Current architecture used for MMOGs based on the classic tightly coupled distributed system. While, MMOGs are getting more interactive same time number of interacting users is increasing, classic implementation architecture may raise scalability and interdependence issues. This requires a loosely coupled service oriented architecture to support evolution in MMOG application. Data flow architecture, Event driven architecture and client server architecture are basic date orchestration approaches used by any service oriented architecture. Real time service is hottest issue for service oriented architecture. The basic requirement of any real time service oriented architecture is to ensure the quality of service. In this paper we have proposed a service oriented architecture for massive multiplayer online game and a specific middleware (based on open source DDS) in MMOG's for fulfilling real time constraints.

## 1. Introduction

Online games give the player the ability to compete against other players over a network [1]. Massively multiplayer online game is a type of online computer game that enables hundreds or thousands of players from various parts of the world to simultaneously interact in a gaming environment they are connected to via the network. Game designers have successfully built multiplayer (MP) and massively multiplayer online games (MMOG) using different approaches. MMOGs were first introduced by various companies as Massive Multiplayer Online Role Playing Game (MMORPG).

A key difference between the multi-player (MP) online game and MMOG suggested in [1] is scale and the associated infrastructure to support it. In MP games, the numbers of concurrent players are between 16 and 32. The game can be played either stand-alone or in multiplayer-network mode, and one of the players machines acts as the server. The game duration is short-lived and if the server crashes, the game is severely disrupted. Today MMOG's, with hundreds of thousands of players online at the same time; also span hundreds of servers. Game session must last for a long time requiring it to be run on dedicated servers equipped with a persistent database. Network bandwidth to support the game-related traffic also comes with a cost. High bandwidth also a means to have better and high quality graphics support. High quality graphics can result in attracting more and more users, thus creating massive user environment. MMOGs have in the past years grown into a million player industry worldwide, especially in Asia, Europe and North America earning massive profit. If more visualization effects and graphics can be added in MMOG, not only more users can be attracted from all over the World but also can lift the profit to another level.

MMOG's are played on network (internet, for example) and computers use protocols to communicate with each other on internet. So choice of protocol is subtle for efficient use of bandwidth. Well known family of protocols belong to TCP/IP family and each protocol solves problems related to its on layer. For the specific requirements of MMOG, it is needed to select a suitable transport layer protocol to provide the exact functionalities required, thus eliminating any overhead that may affect the performance. An example scenario would be like; TCP provide reliable data transfer but can affect the performance due to unnecessary retransmissions in some cases. Choice of protocol for the presented architecture would be described in more detail in one of the upcoming section.

Strict timing constraint for state handling is critical issue too. This is essentially a QoS concern as described in [2]. Very important to a multiplayer game, is the problem of maintaining the same game state information on each of the player's instance of the game and generating the effect of each player belonging to the same game instance. This concern affects the choice of middleware for the architecture. We proposed data distribution standard as middleware having native support to meet this constraint, more details of which would be described in section 4.

The primary contribution of this paper is to propose a loosely coupled paradigm incorporating service oriented architectural concepts and business process execution language for simulation flow. In this paper we discuss a loosely coupled, service oriented architecture mainly to overcome the above mentioned challenges. We have proposed a specific middleware (based on open source DDS) to add all the above said functionalities in MMOG's at real time and modelled simulation flow in business process executed language (BPEL). Most important technical contribution towards design are the service oriented architectural [3] concepts and data centric publish subscribe middleware.

The need for presenting this study arises because current architectures for online gaming applications do not meet scalability requirements and do not provide QoS guarantees. Client-server architecture cannot be assumed to bear load of such a large number of users of MMOG's. Mirrored client-server architecture uses a synchronization technique which is difficult to maintain when large numbers of players are online in MMOG and the environment has become highly dynamic. In addition since each server has a local copy of whole game state, when network becomes highly scalable, additional resources may be needed on the server for brisk processing capabilities. Peer to peer gaming is also less reliable in terms of security as global game state is stored in local peer, hence malicious peers can modify the game state and propagate to other peers.

The rest of the paper is organized as follows: Section 2 presents some related previous work by other authors on MMOG systems while section 3 provides background and preliminaries for real time service oriented architecture. Section 4 presents the architectural details and various implementation scenarios. MMOG specific challenges, requirements and their solutions are described in section 5 and finally section 6 concludes with future work.

## 2. Related work

In this section we discuss current implementation approaches and some pros and cons of the existing MMOG-architectures.

### 2.1. Client-server architecture

Some first person shooter online games like Quake and Doom typically use the client-server architecture. In this architecture a single server is responsible for handling of game states and clients. Quake II [4] like almost all virtual reality games, also follows a popular server based topology in which a single server maintains the state of the game world. The game state is a collection of objects associated with which a small part of the game world. These parts can be computer controlled player, terrain etc. Functions determine the actions of the player and allow freedom to interact with other players. At each iteration, server is responsible for function invocation and assigning players actions. Bharambe et. all in [5] also discusses to some extent the scalability analysis of Quake II. This architecture cannot be assumed to bear load of such a large number of users.

### 2.2. Mirrored & scalable client-server architecture

This architecture is similar to client-server architecture but now each server maintains its own copy of game state. All copies of game states should be identical and there is critical requirement of synchronization between each server. Clients are now distributed amongst a number of servers which reduces the load on a single server. So this architecture presents flexibility in terms of scaling. This technique with the issue of synchronization between servers is discussed in [6]. It is difficult to maintain this synchronization technique when large numbers of players are online in MMOG and the environment has become highly dynamic. In addition since each server has a local copy of whole game state, when network becomes highly scalable, additional resources may be needed on the server for brisk processing capabilities.

### 2.3. Peer to peer games

Another area of extensible research presented some of which in [7], [8] and [9]. Game state is stored in peers and each peer is responsible for its own region. Peer to peer systems are not under the centralized control of the game server. Concept of multicasting is used where client in a periodic and distributed way send update to all other peers involved in a game session. This architecture consumes a lot of bandwidth too. It is also less reliable in terms of security as global game state is stored in local peer. So malicious peers can modify the game state and propagate to other peers. A specific middle-ware is also designed in [9] to work between application and network layer. This layer is specific to peer to peer games and very much application specific considering QoS concerns.

### 2.4. Distributed deployment techniques

Some research designs including specific middle-ware design techniques have used server clusters to improve the scaling of server oriented design. Although server control is desirable for tight administrative policies in some cases, a distributed architecture can address many challenges. As discussed in previous section, scaling is the critical issue for MMOG which can be well addressed having distributed architecture. Single point of failure risk in case of client-server architecture can also be eliminated by distributed architecture. Such design techniques are presented in [10], [11] and [12]. Main idea is to take advantage from locality of interest to distribute the game across several game servers and reduce computational constraints on each single server as well as bandwidth requirement. These designs give ability to handle a very large number of simultaneous users and provide enough computational capability to simulate the gaming algorithms.

Distributed design can make use of third party server deployment also but comes with inter-node communication costs and latencies [12]. These distributed architectures belong to tightly coupled systems making strong assumptions about the interface of interconnected components. So changes in one component's interface reflect it to the entire components interface having significant impacts on all the components. As a result these systems are more difficult to modify and they require a retest of the entire components associated with the same interface. These systems also suffer limitations in independent and incremental development, lack of support to impedance mismatch and no dynamic real time adaptation. In [13] each game server assigns dynamic microcells, each of which contains a very small portion of the large game state. The Microcells can be distributed between servers to balance game load efficiently. It also proposes various methods for efficiently balancing the load. Also microcells depend on a single shared storage mechanism which can be a scalability issue and could potentially suffer from a large performance drawback in MMOG environment.

Various other related techniques have been presented in literatures. In [14] some improvements to microcell management are suggested. It particularly focused on handling game events occurring near virtual boundaries that provided seamless transfer of objects between servers. It also describes an algorithm for dealing with hotspots, and discusses an algorithm that allows the entire game to scale horizontally. Work presented in [15] is more closely related to work presented in this paper. It describes the design and implementation of specific data centric publish subscribe (DCPS) middle-ware for network games. It provides some scalability related features but lacks in reliability concerns. QoS guarantee has not been assured by specific middle-ware. A specific middleware named journey (no open source) has been designed [22] and tested on top of a peer-to-peer network infrastructure to provide load-balancing, fault-tolerance, and cheat-detection capabilities.

## 3. Service oriented architecture

Service-oriented architecture (SOA) provides methods for independent and incremental development and integration of systems. Systems typically assume functionality around business processes and package them as interoperable "services" [16]. SOA also describes an information infrastructure which allows different applications to exchange data with one another as they participate in the related process.

Service Oriented Architecture ensures functionalities like:

- Interoperability among the systems having different software
- Reusability of network resources
- Seamless flow of data from one side to other
- Monitoring and tracking of the information
- Scalability of the already deployed network

This standardized architecture is designed to better support the connection of various systems of systems and the sharing of data. It breaks down large applications into smaller modules as services and unifies different processes. Different groups of people both inside and outside of system can use these applications. Data flow architecture, Event driven architecture and client server architecture are basic date orchestration approaches used by any service oriented architecture [3]. The basic requirement of any real time service oriented architecture is to ensure the quality of service. A simple example of SOA is shown in Fig. 1. Real time service oriented architecture provides real time system operations and interaction between services and provides support to meet with strict timing constraints.

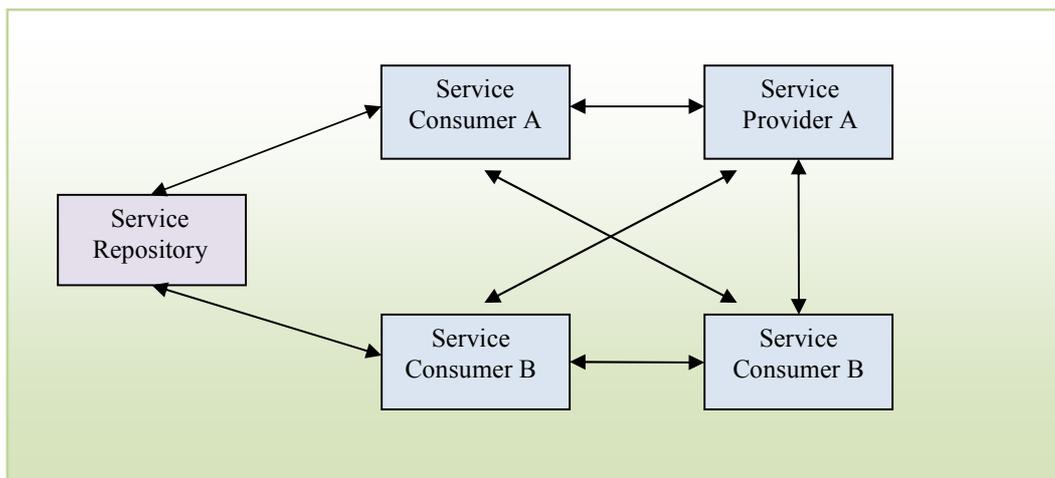

**Figure 1.** Simplified components of SOA

## 4. SOA for MMOG

Based on service oriented concept, conceptual system architecture is presented in Fig. 2.

### 4.1. SQL based database

An SQL based database holds the user attributes such as user login, password, account information and user privilege as shown in table 1.

### 4.2. SQL queries

SQL queries to create tables, populate the database and retrieve records are as follows.

#### 4.2.1. Creating database

CREATE DATABASE game_data;

#### 4.2.2. Creating tables

```
CREATE TABLE user_accounts
 (user_login VARCHAR(25) NOT NULL,
 password VARCHAR(25) NOT NULL,
 user_privillage VARCHAR(25) NOT NULL,
 account_creation_date DATE,
 account_expiration_date DATE);
```

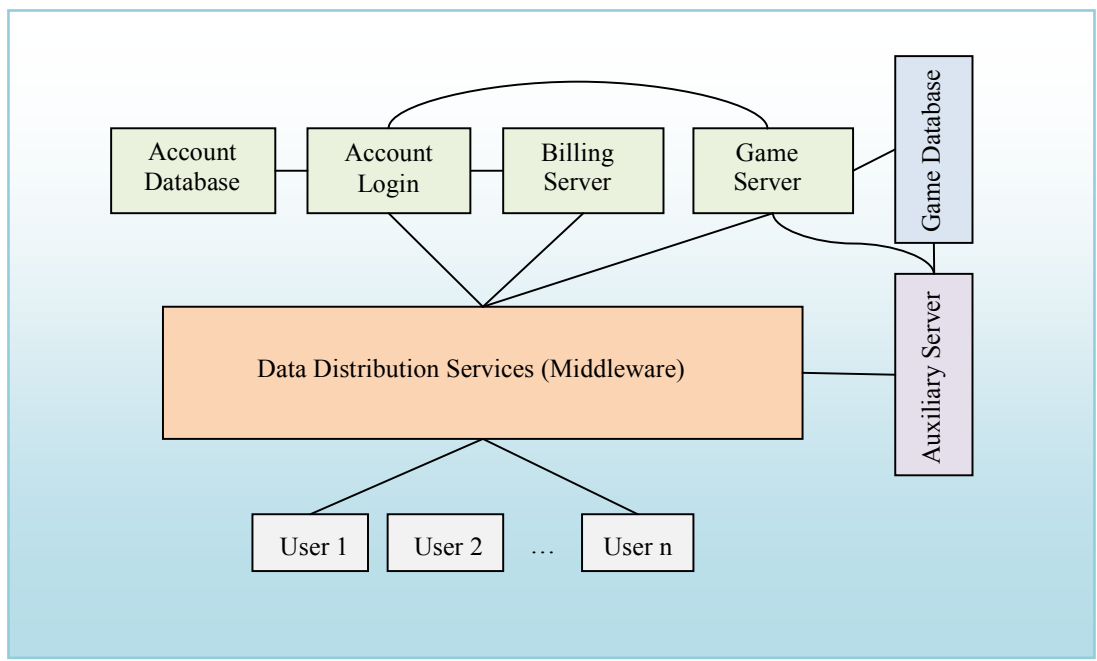

**Figure 2.** Service oriented system architecture for multi-user MMOG

### 4.2.3. Populating tables

INSERT INTO user_accounts ('Max', 'game123', 'Full', '09/10/2008', '09/10/2014');
INSERT INTO user_accounts ('John123', 'helloworld', 'basic', '05/11/2008', '05/11/2013');

### 4.2.4. Retrieving records

SELECT user_privilege, account_expiration_date
FROM user_accounts
WHERE user_login == 'Max'

Table 1. Structure of SQL Based Database

| user_login | password | user_privilege | account_creation_date | account_expiration_date |
|---|---|---|---|---|
| Max | ******* | FULL | 09/10/2008 | 09/10/2014 |
| John123 | ****** | BASIC | 05/11/2008 | 05/11/2013 |
| . | | | | |
| . | | | | |
| . | | | | |

### 4.3. Account login server

This server deals with the information related to the user account. Main purpose of this server is to take login data input from the user side, authenticate user using account database, communicate with the billing and Game servers.

### 4.4. Account billing server

BPEL/XML based server that keep track of user billing information.

### 4.5. Game server

It is a simple online game server that allows user to play single player or multiplayer games.

### 4.6. Auxiliary services server

It can be used to keep track of user's auxiliary actions like chatting. It can also be a part of Game Server.

### 4.7. Simulation of SOA concept for servers

Traditional implementation of these standalone servers is done by simple HTML and PHP scripting with a centralized database. We propose to use a loosely coupled, service oriented implementation for these services as part of MMOG. Advantages achieved through such design components validate the usefulness of our proposed architecture for MMOG [24].

In our presented scenario we have three independent processes that are logically distributed over the network.

    i.    User Authentication Process
   ii.    User Check Process
  iii.    Credit Card Check Process

All the three process have different input output requirements. Processes are implemented using Business Process Execution Language (BPEL). Connection between two different processes is done through logical ports, implemented in Web Service Description Language (WSDL). Such an implementation provides loosely coupled, service oriented scenario.

### 4.8. Simulation flow in BPEL

Business Process Execution Language (BPEL), short for Web Services Business Process Execution Language (WS-BPEL) is an executable language for specifying interactions with Web Services. It has widely become an XML based standard for defining business processes. Business Process Execution Language supports processes which exchange (export and import) information by using Web Service interfaces exclusively. BPEL essentially has a very rich expressive power for describing the behavior of business processes and support service oriented paradigm [17]. Use of BPEL for the composition and implementation of above mentioned process, ensures many functionalities. This provides loose coupling through operations that exchange data only. This differs from component and distributed object models, where behavior can also be exchanged. Operations in these web services based on the exchange of XML data. They are a collection of input, output, and fault messages. The combination of messages defines the type of operation. This differs from previous distributed technologies. It has also support for asynchronous as well as synchronous interactions. It allows information exchange in a stateless manner.

BPEL can be used for real time service oriented architecture (RTSOA) for some business processes where real time requirement is not very tight. The processes which have hard real time requirements cannot be handle by BPEL as it does not have the required capabilities. In our MMOG problem we require more than soft real time which cannot be provided by BPEL. Therefore we have presented a specific middleware to fulfil these requirements.

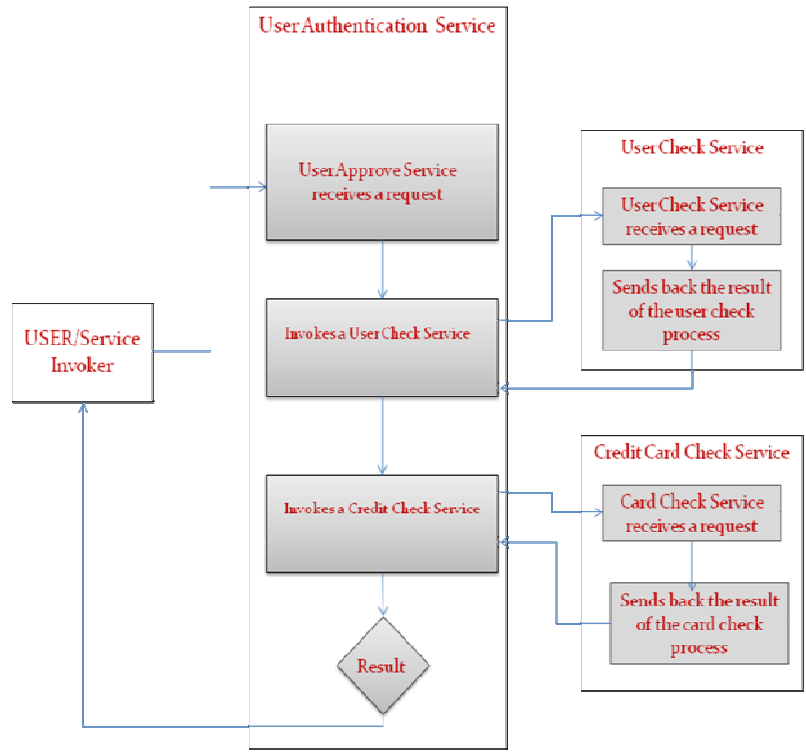

**Figure 3.** Simulation Flow diagram for loosely coupled concept of user authentication block

Fig. 3 describes the flow diagram of all the processes. 'UserApproveService' process invokes two other processes, which performs username authentication and credit card authentication without having dependency upon each other. 'UserCheckService' process is invoked by the main process 'UserApproveService' for authentication. Similarly 'CardCheckProcess' is also invoked by 'UserApproveService' and returns a positive message if approved and negative response if denied. Then 'UserApproveService' makes decisions based on messages provided by other two services. All the processes are deployed over a local GlassFishV2 server.

User Approve Service process invokes two other processes, which performs username authentication and credit card authentication without having dependency upon each other. Also there is no centralized database as was the case with previous implementation with PHP and SQL. UserCheckService process is invoked by the main process 'UserApproveService' for authentication and returns the positive reply if information matches and negative reply if credentials don't match.

### 4.9. Middle-ware for MMOG

For the architecture proposed, middleware is the backbone and choice for this determines the overall reach of the complete architecture. BPEL can prove to be a good choice, as we described in user authentication services. Advantage of using BPEL is a very low deployment cost of the system. Problem with such implementation is the ability of BPEL to support only soft real time processes. As mentioned in previous section, MMOGs may require more than soft real time capability. For a robust solution to this problem we propose to use an open source 'Data Distribution Service (DDS) Standard' from the Object Management Group (OMG), as a middleware. This standard defines an efficient, high-performance publish-subscribe system that offers a predictable way of meeting the data-distribution requirements of data-critical systems with minimal overhead [18].

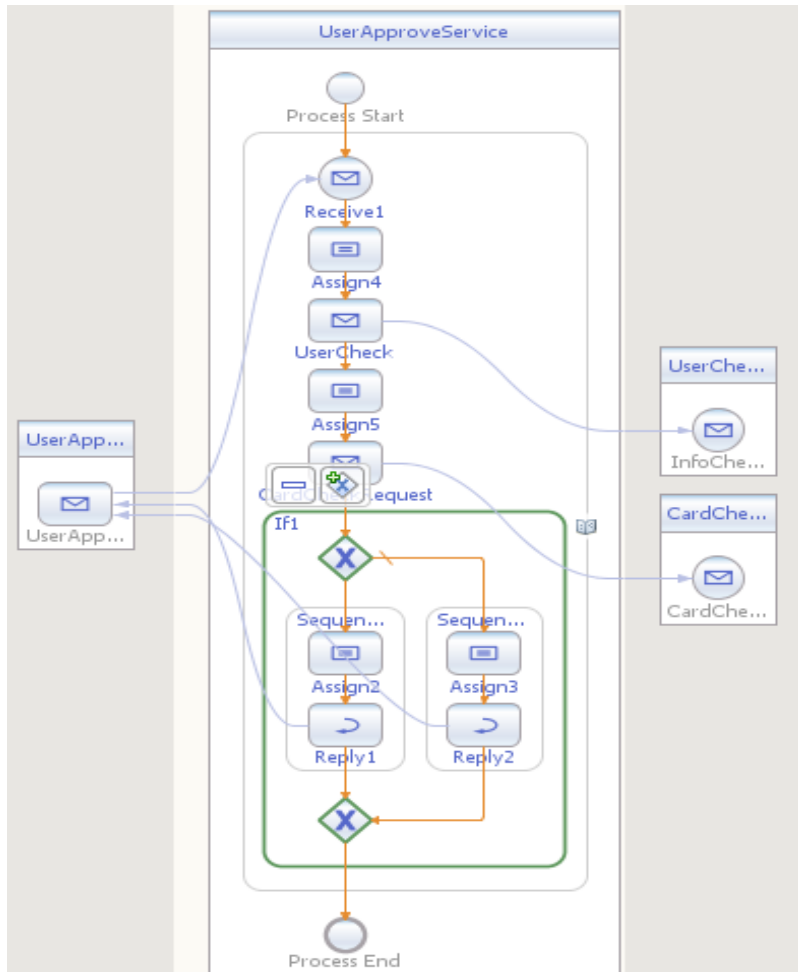

**Figure 4.** Simulation Flow Diagram for User Approve Service

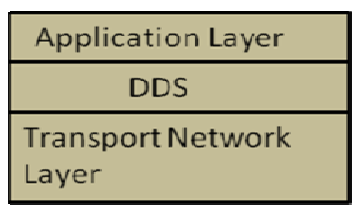

**Figure 5.** DDS standard role in layered architecture

In networked massive multiplayer gaming scenario, due to the scalability limits and single point-of-failure topologies of traditional client/server architectures, DDS specifies an information hub to which game applications at end users can dynamically connect in order to publish and subscribe to information. In contrast to most network games present today, which follows predefined, well implemented protocols like TCP, our gaming architecture must be following the same promises as delivered by DDS middleware as part of real time service oriented architecture. The promises to be delivered are:

   *1) Data Centric Communication Exchange:* Communication network model of game is pure data-centric exchange as required by RTSOA data oriented paradigm. Gaming application, at top layer publishes data, which is then made available to other remote users that are interested in it. In case of at most two players, publisher on one side publishes data and subscriber on other side use it for its application layer and vice versa. This allows communication exchange as pure data centric.

   *2) QoS Policies:* We achieve the required flexibility by this communication exchange. Middleware specifies the available resources and provides policies to ensure the availability of resources to meet the most critical requirements. In this way Quality of Service (QoS) property that affect predictability, overhead, and resource utilization can be controlled. That allows game users to exchange data with good QoS achievable.

   *3) Scalability:* Since scalability as another important issue in multiplayer online games, by building DDS as a middleware in gaming communication network, we promise the system to scale to multiple users having their own publishers and subscribers in a robust manner.

A comparison between existing middleware technologies is given in table 2.

Table 2. Comparing DDS with other Technologies

| Technology | Requirement | | | |
|---|---|---|---|---|
| | Non Real Time | Soft Real Time | Hard Real Time | Extreme Real Time |
| DCOM | Yes | No | No | No |
| JAVA/RMI | Yes | Yes | No | No |
| CORBA | Yes | Yes | Yes | No |
| DDS | Yes | Yes | Yes | Yes |
| MPI | No | No | No | Yes |

## 5. MMOG requirements and solutions

As mentioned in section 4, different standards claim for the provision of RTSOA. Choice of data distribution standard is justified since not only it can handle RTSOA but can also be used to address many issues specific to MMOGs. In the following we discuss the additional support provided by the middleware system to address common issues that must be resolved in order to achieve a realistic MMOG behaviour.

One of the major issues associated with MMOG is its scalability and communication performance that enables the interactive system to maintain the quality and coordination of the game experience. Scalability means, the ability of a system to respond to an increasing demand from its users, without significantly degrading the quality of the interactive experience. This is overall an architecture that enables the integration of new components, resources and technologies for system expandability. This architecture is service oriented in nature and it maintains loosely coupled entities. In loosely coupled paradigm all entities are independent and support scalability. The quality and coordination of communication performance is achieved by QOS block of Data Distribution Service. MMOGs are typically migrating towards 3D environment for player's interaction. These environments can become quiet dynamic, making it impossible for the clients to save the virtual world state hoping that when the player returns the game state remains the same. Presence of history policy as a part of QOS Block in this architecture can address issues related to state handling of the game

MMOG requires different multimedia (Audio, video, text) objects; efficient temporal presentation of these is an important question. Proposed middleware can handle issues related to presentation of multimedia objects by using presentation policy in the QOS Block. For a single instant of game when having multiple users (a group of users), there arises a need for a policy to handle group based data. To handle group based data the middleware has in built Group Data policy in QOS block [18]. Group data policy along with Data Reader Listener and Data Writer Listener provide policies to manage group data. If we have multiple changes in some object by different users than we need to send the coherent changes to appropriate users; this can be managed in presentation policy provided by DDS. To meet the requirement of reliable data service in MMOG, reliability policy present in QOS block of the middleware can be implemented. MMOG requires smooth playback even in the presence of network latency or delay. Network Latency should not block the game from updating for an unnecessary long time. It is highly undesirable to continue the game after a reply message arrives. Suspend and Resume publication policies in a publisher block used by DDS [18], can handle this situation.

In MMOG different users can join or leave the game. A good middleware should detect the arrival or departure of clients. In DDS we have Delete publisher, Create Publisher, Delete Subscriber and Create Subscriber policies to handle the issue of arrival and departure of users. In MMOG virtual environment can become quite big, and a significant part of it is not relevant to the action context of a player at a certain time. Therefore each player does not need to receive notice of all the changes that are happening everywhere else in the virtual environment. Support for such kind of scenario can be provided by event filtering policy to subscribe only to the data which is relevant to a particular user.

The problem of maintaining the same game state information on each of the player's instance of the game and generating the effect to other players is quite subtle. For example in a shooting game (likewise in racing game) the exact positions of all the player's characters must be represented on the player's screens at the same time. If there was a delay in the updating of the position of the characters then two players may have wrong game state displayed on their screens which in turn may end with wrong results not in accordance with player's actions. Similarly in racing game this may result in a scenario that both player's car are at first position as perceived by players. In the middleware QoS Policy block have 'Listener' and 'Status' interfaces. 'Listener' provide a mechanism for the service to asynchronously inform the application of relevant changes in the communication status and 'Status' represents the communication status of the players. Change in status values asynchronously informs the application through these interfaces.

## 6. Pros and cons of MMOGs

Because of the growing broadband network, multi-player online games can be played while interacting with the other players and without meeting them. Here are some pros and cons of multi-player online games. Online games provide privacy. We need not to reveal the true identity to other users while maintaining interaction with them. Online games are cheaper because it can be played from the comfort of our time playing for as long as we want.

Massive multi-player online games are not early to be finished and these are time consuming. These games are quite competitive and the players need time to think of strategies to accomplish their goals. In general these role playing games lessen social interaction. Because of the large amount of time players spend, interacting with other players mostly through chat, they become impersonal in dealing with people in real life.

## 7. Limitations and open challenges

SOA for MMOG will require a large investment by way of technology, development and trained staff. We discuss the limitations of SOA deployment below. Some of the generalized issues have also been discussed in [19].

1) Since services described in section 4, invoke other services, each service needs to validate completely every input parameter. This may have negative impact on response time.
2) Any malicious activity introduced in a well-used service may take out the entire application, at some times.
3) When using data distribution service (DDS), an open standard middleware, a service is much more open to other services and applications and thus security becomes an issue.
4) Service management becomes an issue since SOA blurs the boundaries of application ownership. It has been recommended in [21] that a multi-gaming-vendor solution can fully benefit from the flexible nature of SOA.

Open source implementation of the Data Distribution Standard as middleware is still a technical challenge. Hierarchical design of the standard will require proper integration with the application and specific algorithms for various functionalities which needs to be explored as an open research issue. Current work also needs to be extended to increase the flexibility of resource allocation. In an MMOG cloud environment, virtual machines (VMs) in game servers are allocated to serve players [23]. SOA based MMOGs can also be integrated with cloud computing architectures for efficient resource allocation and a cost effective solution.

## 8. Conclusion

MMOG architecture require much more than classic tightly coupled distributed services. In this paper we have proposed a service oriented architecture for the deployment of MMOGs. Such architecture provides loosely coupled distributed services. Taking it a step further, we also considered real time requirements for such applications. Our proposed architecture achieves much of its capabilities from an underline middleware and ensures both service oriented and real time requirements. For the scheme presented in this paper we considered DDS standard as middleware. DDS standard has shown success in various proprietary applications having critical RTSO requirements. To demonstrate the capabilities of our proposed loosely coupled service oriented architecture, we implemented a prototype using business process execution language. The ease with which a new process can integrate in the implemented prototype and level of independence available to the processes, show the usefulness of such architecture.

## 9. Future work

Effectiveness of the proposed architecture highly depends on the choice of middleware. Open source implementation of the Data Distribution Standard as middleware is not yet available. Hierarchical design of the standard will require proper integration with the application and specific algorithms for various functionalities. Further, software models [19] using service oriented concepts for MMOG can be developed and implemented as a software package. Effective resource allocation and cost effective solutions as well as scheduling algorithms for anycast and multicast routing [20] on underlying network layer protocols need to be investigated.